# Elastic Compliance and Stiffness Matrix of the FCC Lennard-Jones Thin Films: Influence of Thickness and Temperature

Joël Puibasset*

Interfaces, confinement, Matériaux et Nanostructures, UMR 7374, CNRS, Université d'Orléans, 1b rue de la Férollerie, 45071 Orléans cedex 02, France

*e-mail : puibasset@cnrs-orleans.fr

**Abstract**

The fcc Lennard-Jones crystal is used as a generic model of solid to study the elastic properties of thin films as a function of thickness and temperature. The Monte Carlo algorithm is used to calculate the average deformations along the axes in the isostress-isothermal ensemble that mimics a real uniaxial loading experiment. The four independent parameters (tetragonal symmetry without shear) have been calculated for film thicknesses ranging from 4 to 12 atomic layers, and for five reduced temperatures between 0 and 0.5 $\varepsilon/k_B$, where $\varepsilon$ is the energetic parameter of the Lennard Jones potential and $k_B$ is Boltzmann's constant. These parameters (Poisson's ratio and moduli) give the compliance matrix, which is inverted to get the stiffness coefficients. It is shown that the three Poisson's ratios exhibit a good linearity with the inverse of the film thickness, while this is not the case for the moduli and the compliance coefficients. Remarkably, the stiffness coefficients do exhibit a good linearity with the inverse of the film thickness, including the limiting value of infinite thickness (bulk solid) obtained by applying periodic boundary conditions in all directions. This linearity suggests to interpret the results in terms of a bulk+surface decomposition. However, the surface stiffness matrix deduced from the slopes has nonzero components along the out-of-plane direction, an unexpected observation in the framework of the surface stress theory.

# I Introduction

Solids are not rigid systems: their deformation under loading[1] can play an important role in various situations. In particular, in the context of fluid adsorption in solids, it has long been observed that the presence of a condensable gas in a porous material induces its deformation.[2,3] Obviously, soft or deformable systems exhibit large deformations which are due to and reciprocally influence adsorption.[4-7] But in most cases, adsorbents are solid bodies, and the deformations are small (elastic regime).[8-12] The theory of poroelasticity has been developed in this context to study the adsorption-induced deformation of systems like soils or reservoirs, where it can play an important role.[13-19] The smaller the pores, the larger the capillary forces that induce the deformations: it is thus important to have good models of the mechanical properties of nanometric systems in order to be able to predict their deformation during adsorption. Conversely, the deformation of a saturated nanoporous medium along the main desorption curve is also a powerful tool to measure the elastic constants of nanometric systems, in complement to direct measurements.[20-32] The knowledge of the fluid-solid interactions has also proven to be important due to specific effects.[33]

Understanding the mechanical properties of nanosystems is also important in the context of nanoporous solids, in particular regarding the adsorption/desorption hysteresis that appears for a sufficiently low temperature. Many explanations have been proposed: the pore morphology which influences the adsorption and desorption mechanism, the topological effect of the porous network, the disorder created by the solid adsorbent, etc.[34-49] The observation of correlations between the adsorption/desorption hysteresis and the elastic deformations[50-53] has been the starting point of recent studies which suggest that the free energy variations due to the solid deformations could affect the isotherm, and explain the shape of the hysteresis curve, in particular the sharp desorption branch generally associated with collective or avalanche effects.[54-57]

Very recently, the development of nanometric devices (nanoelectromechanical systems, NEMS)[58] has been the starting point of many studies, in particular on silicon nanocantilevers.[59-62] It is thus desirable to have accurate models of the elastic properties of nanometric systems, and many studies have focused on the size dependence of the mechanical properties of nanometric devices. The effect of thickness is generally attributed to the surface stress induced by the free surfaces.[63-72] A clear understanding of experimental data is still lacking, and theoretical approaches have been developed for a long time.[73-80] The nanometric size being also well-suited for atomistic studies, many algorithms have been proposed to track the problem by molecular simulations in various statistical ensembles.[81-89] These molecular simulations are useful for accurate predictions since in principle any well-suited interatomic potential may be used to describe the interactions.

On the other hand, non-specific mechanical properties of solid may be studied using simple pairwise additive interactions like the Lennard Jones potential.[90-94] These simple models show that surface relaxations occur over few atomic layers: the effects are thus enhanced in ultrathin films or multilayers.[65-67,95-105] The simple decomposition of the elastic properties in a bulk + surface contribution, and the expected linear dependence with the inverse system size, remain controversial.[64,66,88,97,106,107] Furthermore, the effect of temperature is not so frequently studied.[83,108] The following study thus focuses on a simple Lennard-Jones thin film model to determine the effect of both the film thickness and the temperature on its elastic properties. We present the model and methods, in particular the two numerical deformation experiments used to calculate the elastic constants of the thin film. The results are then given for various film thicknesses and temperatures, and discussed in terms of surface stress.

**II Model and methods**

We investigate the elastic properties of atomic thin films, in particular the influence of temperature and film thickness. The model is the Lennard-Jones (12-6) crystal that catches the main features of elastic solids. The approach can be easily extended to any solid of interest described with more accurate potentials. The procedure used to determine the elastic constants follows the experimental route. In this computational method the elastic constants are defined as the initial slope (linear regime) of stress-strain curves given by the molecular simulations.[82,83] The calculations are limited to deformations without shear, and the statistical averages are performed in the framework of the Monte Carlo simulations, well-suited to monitor strain fluctuations in the constant pressure ensemble.[89]

We have first determined the bulk elastic constants as a function of temperature. Then, the system is cut along one crystallographic plane and the corresponding periodic boundary conditions are removed, in order to obtain a film with two free surfaces. The film thickness is varied gradually by removing the external layers, and the film elastic constants are determined for each thickness. The reduced symmetry of the film compared with that of the bulk solid has consequences on the number of elastic constants to be determined and on the method to measure these parameters.

**II_1 Stiffness and compliance matrices**

The stress tensor is the derivative of the free energy of the solid with respect to the strain tensor components. In the limit of small deformations, the stress tensor is linear with the strain tensor (Hooke's law): the elastic stiffness matrix $C$ is defined as the fourth rank tensor that relates stress to strain. The Voigt notation will be used to simplify the equations. The generalized Hooke's law then writes:[1]

$$\sigma_i = \sum_{j=1}^{6} C_{ij} \varepsilon_j \qquad (1)$$

with $\sigma_1 = \sigma_{xx}$, $\sigma_2 = \sigma_{yy}$, $\sigma_3 = \sigma_{zz}$, $\sigma_4 = \sigma_{yz}$, $\sigma_5 = \sigma_{xz}$, $\sigma_6 = \sigma_{xy}$, $\varepsilon_1 = \varepsilon_{xx}$, $\varepsilon_2 = \varepsilon_{yy}$, $\varepsilon_3 = \varepsilon_{zz}$, $\varepsilon_4 = 2\varepsilon_{yz}$, $\varepsilon_5 = 2\varepsilon_{xz}$, $\varepsilon_6 = 2\varepsilon_{xy}$, where $\sigma_{\alpha\beta}$ and $\varepsilon_{\alpha\beta}$ are the stress and strain tensors respectively, for which $\alpha$ and $\beta$ equal $x$, $y$ or $z$; $C_{ij}$ is the stiffness matrix.

The stiffness matrix is symmetric. Furthermore, the intrinsic symmetry of the solid reduces the number of independent coefficients in the matrix. Choosing [100], [010] and [001] as the axes, the cubic symmetry of the bulk fcc crystal imposes that only three independent elastic constants are expected ($C_{11} = C_{22} = C_{33}$, $C_{12} = C_{13} = C_{23}$, and $C_{44}$). For a film of infinite extension in directions x and y, and cut from the bulk crystal along parallel planes perpendicular to the z axis (see Figure 1), the tetragonal symmetry implies that the compliance matrix has only seven distinct coefficients. Since there is no shear stress in the mechanical tests that we have performed, the shear components will be discarded. The stress-strain relation then reduces to:

$$\begin{pmatrix} \sigma_1 \\ \sigma_2 \\ \sigma_3 \end{pmatrix} = \begin{pmatrix} C_{11} & C_{12} & C_{13} \\ C_{12} & C_{11} & C_{13} \\ C_{13} & C_{13} & C_{33} \end{pmatrix} \begin{pmatrix} \varepsilon_1 \\ \varepsilon_2 \\ \varepsilon_3 \end{pmatrix} \qquad (2)$$

with only four independent coefficients. The stiffness coefficients $C_{ij}$ may be obtained directly from imposed deformations applied to the system along the three axes, and measuring the corresponding components of the stress tensor (finite-strain method). Conversely, constant pressure or stress may be applied to the system: the average deformation in each direction then gives the compliance constants. When expressed in terms of Young's modulus ($E$) and Poisson's ratio ($\nu$), the compliance matrix of the film writes:

$$\begin{pmatrix} \varepsilon_1 \\ \varepsilon_2 \\ \varepsilon_3 \end{pmatrix} = \begin{pmatrix} \frac{1}{E_1} & -\frac{\nu_{21}}{E_1} & -\frac{\nu_{31}}{E_3} \\ -\frac{\nu_{12}}{E_1} & \frac{1}{E_1} & -\frac{\nu_{31}}{E_3} \\ -\frac{\nu_{13}}{E_1} & -\frac{\nu_{13}}{E_1} & \frac{1}{E_3} \end{pmatrix} \begin{pmatrix} \sigma_1 \\ \sigma_2 \\ \sigma_3 \end{pmatrix} \qquad (3)$$

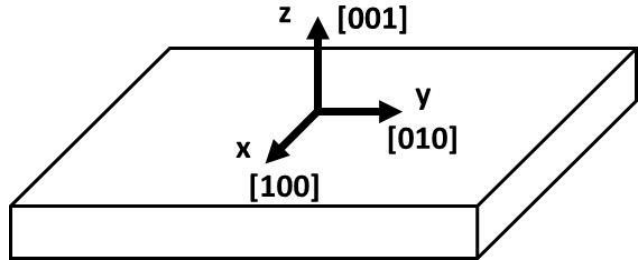

**Figure** 1: Schematic representation of the thin film showing the axes parallel to the crystallographic directions.

The two x and y axes parallel to the film being equivalent, $v_{21} = v_{12}$, as expected from the symmetry of the matrix. On the other hand, $v_{31} \neq v_{13}$, but the symmetry of the matrix imposes that:

$$\frac{v_{31}}{E_3} = \frac{v_{13}}{E_1}. \qquad (4)$$

We thus have 5 parameters and a relation, which gives only four independent parameters as for the stiffness matrix (eq 2). In the case of the bulk solid, $E_3 = E_1$ and $v_{13} = v_{12}$, reducing the number of independent parameters to two.

**II_2 Imposed stress method.**

The coefficients of the compliance matrix are obtained by applying uniaxial loading and measuring the average deformation.

*For the bulk crystal*, the three axes being equivalent, the two independent parameters are obtained in a single simulation: the stress is applied along the x direction ($\sigma_1$, 0, 0), and the measured deformation along the same axis gives Young's modulus $E_1 = \sigma_1 / \varepsilon_1$, while the measured deformation along one of the two other equivalent axes gives the Poisson's ratio $\nu_{12} = -\varepsilon_2/\varepsilon_1 = -\varepsilon_3/\varepsilon_1$.

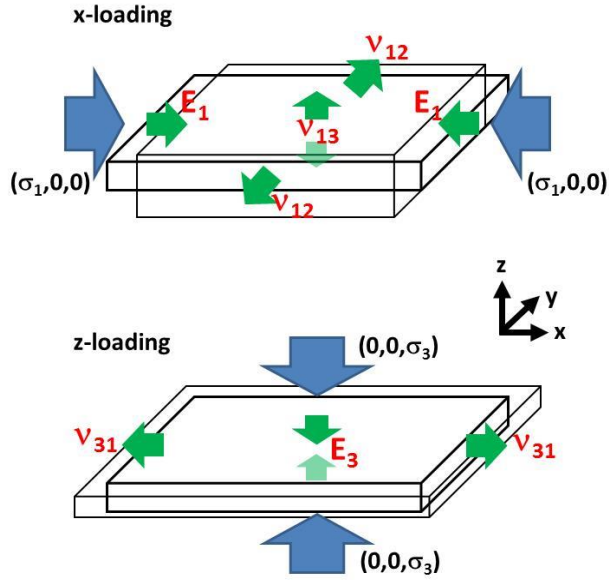

**Figure** 2: Schematic representation of the x- and z-loading experiments. The uniaxial stress is applied along the x or z directions (in blue). The solid deformation is measured along the axis (in green, 5 measurements) which allow to calculate the two Young's modulus and three Poisson's ratio. The redundancy is used to check the symmetry of the compliance matrix.

*For the thin film*, at least two simulations are required (see Figure 2). (i) In the first one, a uniaxial stress is applied along the x axis, parallel to the film, ($\sigma_1$, 0, 0). The measured deformations along the x, y and z directions are all informative, and give respectively the Young's modulus $E_1 = \sigma_1/\varepsilon_1$, and the two Poisson's ratio $\nu_{12} = -\varepsilon_2/\varepsilon_1$ and $\nu_{13} = -\varepsilon_3/\varepsilon_1$. (ii) The second simulation consists in applying the uniaxial stress along the z direction, perpendicular to the surface, (0, 0, $\sigma_3$). The measured deformation along the same axis gives the second Young's modulus $E_3 = \sigma_3/\varepsilon_3$. In principle, at this point, all parameters are determined. However, it will prove useful to calculate independently the third Poisson's ratio $\nu_{31}$ by measuring the deformation along one of the two equivalent directions parallel to the film: $\nu_{31} = -\varepsilon_1/\varepsilon_3$. This will allow to cross-check for the consistency of the five independent measurements through the symmetry relation (eq 4). We thus introduce the quantity $\alpha = \dfrac{E_1 \nu_{31}}{E_3 \nu_{13}}$ which is expected to be equal to 1 for a symmetric matrix.

**II_3 Atomic Model**

The numerical bulk samples were prepared by positioning atoms on the site of an fcc regular lattice. The initial lattice parameter is chosen so that nearest neighbor atoms are at a distance corresponding to the van der Waals diameter. For the bulk measurements, two simulation box sizes have been chosen, equal to 6 and 7 lattice units along each direction (the axes are chosen parallel to the crystallographic directions). The number of atoms is respectively 864 and 1372. Periodic boundary conditions are applied in the three directions. The two systems are relaxed at zero temperature and zero pressure (Monte Carlo run in the isostress-isothermal ensemble). These relaxed systems are then used to calculate the bulk elastic constants as a function of temperature.

The thin films are issued from the largest bulk box. A gap is introduced in the z-direction so that the opposite faces of the film do not interact through the periodic boundary conditions (see Figure 1). The film thickness is varied by removing atomic planes on both sides. Five thicknesses are considered, equal to 2, 3, 4, 5 and 6 lattice units (corresponding to 4, 6, 8, 10 and 12 atomic planes).

All atoms interact with the Lennard-Jones potential:

$$V_{LJ}(r) = 4\varepsilon\left[\left(\frac{\sigma}{r}\right)^{12} - \left(\frac{\sigma}{r}\right)^{6}\right]. \tag{5}$$

where $\varepsilon$ represents the binding energy between atoms and $\sigma$ their van der Waals diameter. A cutoff radius of $4\sigma$ is used, and a shift of the potential at the cutoff radius is applied to avoid discontinuities. For comparison, Quesnel et al.[83,92] have used the same potential with a cutoff radius of $3\sigma$. Small differences are expected, because of the importance of the long range contributions to the interactions, as shown by Tretiakov and Wojciechowski.[79] Their work shows that using $4\sigma$ instead of $3\sigma$ significantly improves the accuracy, the discrepancy with the full potential including the long range corrections being less than few percent. The Lennard-Jones parameters will be used as fundamental units in which all other quantities can be expressed (reduced units): distances are given in units of $\sigma$, temperature is given in units of $\varepsilon/k_B$, where $k_B$ is Boltzmann's constant, and pressure and moduli are given in units of $\varepsilon/\sigma^3$. For instance, the reduced dimension of the initial lattice unit is $\sqrt{2}\sigma$. Numerical values in SI units for quantities calculated with this potential can easily be obtained from the Lennard-Jones parameters tabulated for a number of materials.

**II_4 Simulation method**

The calculations of the elastic constants are performed at five temperatures between 0 and 0.5 $\varepsilon/k_B$. The zero-temperature calculations are done by minimization of the energy of the system, while the finite temperature calculations are done with the Monte Carlo algorithm to sample the system phase space. Clavier and collaborators have shown that Monte Carlo is the best approach for constant pressure runs.[89] Since the imposed thermodynamic parameters are the temperature and the stress or pressure, we use the isostress-isothermal ensemble. The Monte Carlo trials consist in particle displacements to thermalize the system, and homothetic dilatations or contractions along the three space directions to equilibrate the system with the three components of the external stress $\sigma_i$. Different configurations have to be considered for the bulk and the film calculations.

*For bulk calculations*, the simulation box is free to deform in all directions independently. The stress is uniaxial, *i.e.* a finite pressure is applied in one direction (chosen to be x), while zero pressure is applied in the y and z directions. The solid deformation in a given direction is given directly by averaging the fluctuating dimension of the simulation box along that direction. As an example, Figure 3-a shows the natural fluctuations of the box size along y, denoted Ly, at $T = 0.2\,\varepsilon/k_B$, and for the bulk sample with six unit cells in all directions. Similar fluctuations (not shown) are observed for the other free dimension $L_z$, as expected from the symmetry of this uniaxial test along x. The fluctuations follow a Gaussian distribution of width $0.018\,\sigma$ and average $9.33\,\sigma$ shown in Figure 3-c (blue curve).

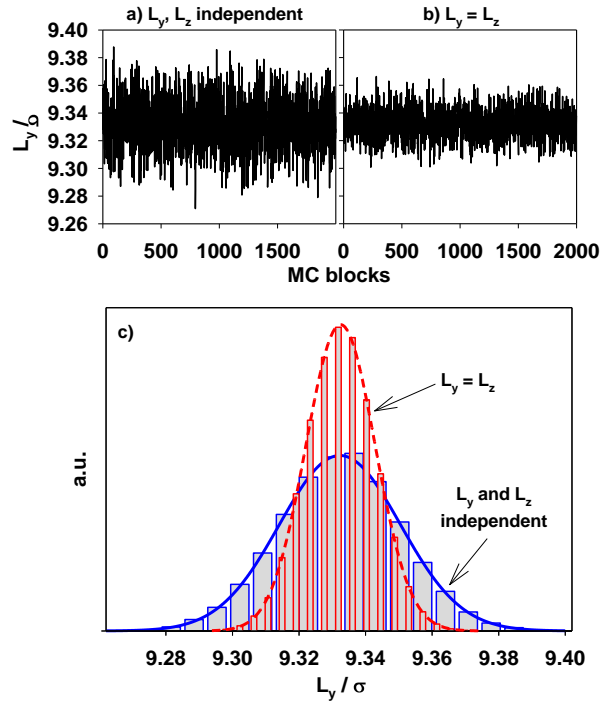

**Figure** 3: a) Spontaneous fluctuations of the lateral dimension $L_y$ of the bulk sample containing 864 atoms (6 unit cells in each direction), during a uniaxial loading along x ($\sigma_1$, 0, 0) at $T = 0.2\,\varepsilon/k_B$. Similar fluctuations are expected for the equivalent z direction (not shown). b) same as a), but with the geometrical constraint $L_y = L_z$ during the course of the simulation). c) Distribution of the fluctuations in both cases (bins) and the corresponding Gaussian fits (lines).

*For film calculations*, the simulation box is free to deform only in the two directions parallel to the film (x and y), while the box size is fixed along the z direction: this is necessary to avoid the shrinkage of the simulation box along z. The stress is applied in directions x and y using the conventional isostress algorithm, and the corresponding solid deformations are given by the box size. One has to be careful in choosing the value of the stress parameters for the simulation code. In the Monte Carlo algorithm, the stress is applied to the whole lateral faces of the simulation box (for example $L_y \times L_z$ for the ($\sigma_1$, 0 0) uniaxial loading). But in a real experiment, it is applied only to the (smaller) lateral surface of the film: $L_z$ is replaced by the film thickness. The value of the stress parameter is thus corrected so that the total force applied laterally on the film is the same in the simulation and in the real experiment. This applies only for the stresses in the x and y directions. In the z direction, the pressure is applied directly and uniformly to the atoms located on the outermost atomic planes. The force acting on each atom equals $\sigma_3 L_x L_y / N$ where *N* is the number of atoms in each plane. Note that the deformation parallel to the film is so small that the forces (proportional to $L_x L_y$) are essentially constant during the simulation. The deformation of the solid along z is monitored by measuring the average position $z_\pm$ of the outermost planes: the film thickness is then given by $z_+ - z_-$.

*The Gibbs' surface:* Some subtleties have to be carefully considered in the case of the film. In a real experiment, the pressure would be applied to the film by another solid, or a fluid. The thermodynamic analysis of the energy received by the system during a mechanical test requires the introduction of the Gibbs' dividing surface situated at mid-distance between the film and the solid or the fluid used by the experimentalist to apply the pressure. If the film deforms ($\varepsilon_3$) the work received equals $\sigma_3 L_x L_y H_z \varepsilon_3$ where $H_z$ is the distance between the two Gibbs' surfaces. In our simulations, the forces are applied directly to the center of the outermost atoms, and the corresponding work received by the solid is $\sigma_3 L_x L_y h_z \varepsilon_3$ where $h_z = z_+ - z_-$ is the distance between the outermost atomic planes. The difference between $H_z$ and $h_z$ is typically one van der Waals diameter: the algorithm thus underestimates the energy transferred to the solid, and thus produces incorrect deformations. Another way to see the same problem is to notice that the deformation $\varepsilon_3$ has to be calculated based on the variations of $H_z$ and not $h_z$, which is equivalent to measuring it in the core of the solid. For thin films, monitoring the total thickness $h_z$ improves the statistics. We thus have to apply either the correction factor $H_z / h_z$ to the stress or its inverse to the strain. Without this correction the compliance matrix is not symmetric, the discrepancy increasing for very thin films. In our simulations, the measured value of the parameter $\alpha = \dfrac{E_1 \nu_{31}}{E_3 \nu_{13}}$ introduced to measure this symmetry decreases from 0.9 for the thickest film down to 0.75 for the thinnest one! On the other hand, with the Gibbs' surface correction, the $\alpha$ parameter oscillates between 0.996 and 1.02: the symmetry of the matrix is well verified.

*Improving accuracy:* The errors on the average values of deformation can be evaluated from the standard block analysis.[109] The typical relative error that can be reached is within $4 \times 10^{-5}$ for the lateral dimensions of the solid, for $2 \times 10^6$ Monte Carlo steps per atom. The situation may be improved if one takes into account the natural symmetry of the system. For instance, in the bulk case (x-loading) or for the thin film (z-loading), one expects identical box deformations (in average) in the y and z directions for the bulk, or the x and y directions for the film. The method then consists in imposing the geometrical constraint $L_y = L_z$ (for the bulk case) during the course of the simulation. The fluctuating value of $L_y = L_z$ is displayed in Figure 3-b. As can be seen the average value of $L_y$ is not affected, but the fluctuations are smaller (the geometrical constraint stiffens the system). The same Monte Carlo run then results is more accurate results. The corresponding relative error is now within $2\times10^{-5}$. This procedure will be used each time it is possible: for the bulk calculations (uniaxial loading) and for the film calculations during the z-loading experiment (Figure 1-b).

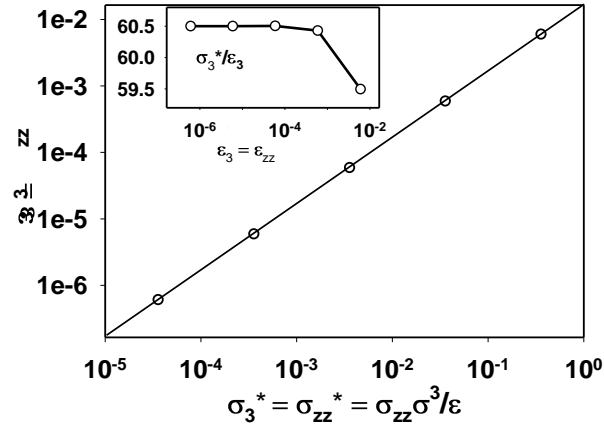

**Figure** 4: Main circles: strain-stress data at zero temperature during the uniaxial loading along the z axis, for an fcc Lennard-Jones solid described by Eq 5. The box dimensions are six unit cells in the three directions (the simulation box is parallel to the crystallographic axes), and periodic boundary conditions are applied in all directions. Calculations are done by energy minimization in the isostress ensemble. The reduced stress ($\sigma_3^* = \sigma_3 \sigma^3/\varepsilon$) and strain vary over four orders of magnitude (logarithmic scales). Solid line: first bisector of the $(\sigma_3^*, \varepsilon_3)$ plane, as given by a linear stress-strain relation. Inset: Effective reduced modulus $\sigma_3^*/\varepsilon_3$, as a function of the strain. The reduced Young's modulus $E_3^*$ is obtained as the limiting value of this quantity for small deformations (linear regime).

**III Results and Discussion**

**III_1 bulk Lennard-Jones crystal**

**Zero temperature calculations:** The first calculations have been performed at zero temperature for the bulk solid for the two system sizes. The aim was to determine the best choice for the stress amplitude in order to be in the linear regime of small deformations. Figure 4 gives the results of the deformation $\varepsilon_3 = \varepsilon_{zz}$ as a function of the applied uniaxial stress $\sigma_3 = \sigma_{zz}$, that has been varied over four decades. The behavior is very close to linearity. The effective modulus $\sigma_3/\varepsilon_3$ has been calculated and shown in the inset (in reduced units). By definition, the Young's modulus $E_3$ is the limiting value of this quantity for infinitesimal deformations. As can be seen, this quantity is not constant for large deformations ($>10^{-3}$), but is essentially constant for small deformations ($<10^{-3}$), as expected in the linear regime. At zero temperature, there is no limitation is choosing the deformation as small as possible except that due to the finite numerical precision: we get $E_3 = 60.5\,\sigma^3/\varepsilon$. The Poisson's ratio is deduced from the deformation in the direction perpendicular to the uniaxial stress: $\nu_{12} = -\varepsilon_2/\varepsilon_1 = 0.364$. These results are comparable with those obtained by Quesnel *et al.*[83,92] who have used a slightly different cutoff ($3\sigma$) and the constant-rate deformation method ($E_3 = 61.1\,\sigma^3/\varepsilon$, $\nu_{12} = 0.361$ for the initial value in the first 0.2% strain).

In order to evaluate the system size effects, the same calculations have been done in the larger cubic box comprising seven unit cells in each direction. The measured Young's modulus at zero temperature is $E_3 = 61.5\,\sigma^3/\varepsilon$ and the Poisson's ratio remains equal to 0.364. The system size effects are small (less than 2%). This has to be correlated with the fact that the simulation box is at least twice as large as the cutoff radius of the potential. These values are in good agreement with the work of Tretiakov and Wojciechowski[79] who take into account the long range interactions and obtain $E_3 = 61.4\,\sigma^3/\varepsilon$ and $\nu_{12} = 0.364$.

**Finite temperature calculations:** The reduced temperature $T^* = k_B T / \varepsilon$ is varied between 0.1 and 0.5. The system size now fluctuates, and the strain-stress curves are obtained from the average values of the deformation. As previously, the calculations have been performed on several decades in order to find the best compromise between linearity (obtained for the smallest deformations) and numerical accuracy (obtained for the largest deformations). The error bars have been deduced from the standard block analysis. In practice, the higher the temperature, the larger the fluctuations. At $T^* = 0.1$, the fluctuations are already so large that deformations smaller than $10^{-5}$ cannot be measured, *i.e.* the error bars on the average deformation exceeds its value. For larger, measurable deformations, the corresponding effective Young's modulus $E_3 = \sigma_3/\varepsilon_3$ is given in Figure 5 as a function of the applied stress, and for all temperatures. As can be seen, the lowest stress values (below $10^{-2}$) are now inappropriate at finite temperature because of the large uncertainty due to natural fluctuations of the system. On the other side, for a reduced stress larger than 0.1, the effective Young's modulus shows small variations: the system is entering the nonlinear regime of deformations. The best range of stress values is between $10^{-2}$ and $10^{-1}$, corresponding to deformations around $10^{-3}$. The calculated elastic constants for the bulk crystal as a function of temperature are given in Table 1 and Figure 6. The data exhibit a global decrease of the modulus with increasing temperature, and a constant Poisson's ratio within errors, in agreement with the work of different authors.[79,83,89] Note that despite small error bars, the Young's modulus data look scattered around the expected linear behavior: this most probably reveals systematic errors missed by the block analysis. In particular, large time scale fluctuations inaccessible to our simulations could contribute to the results.

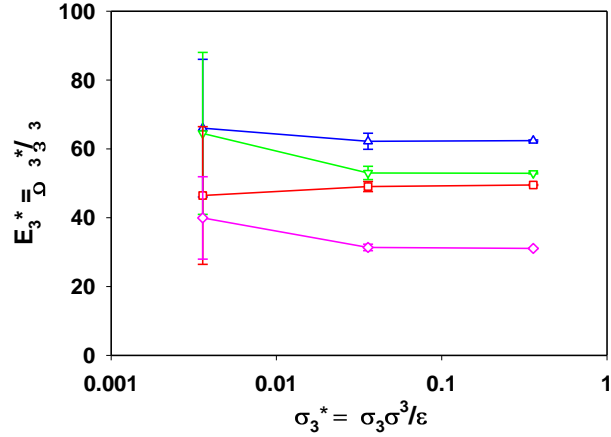

**Figure** 5: Effective reduced modulus $\sigma_3^*/\varepsilon_3$ as a function of the reduced stress $\sigma_3^*$, for the bulk fcc Lennard-Jones crystal at different reduced temperatures $T^* = k_\text{B}T/\varepsilon$ (up triangles: 0.1; down triangles: 0.2; squares: 0.3 and diamonds: 0.5). In the linear regime of small deformations, this effective modulus is expected to reach a constant value (the Young's modulus), but numerical errors become important. The best compromise between accuracy and linearity is for the intermediate stress values.

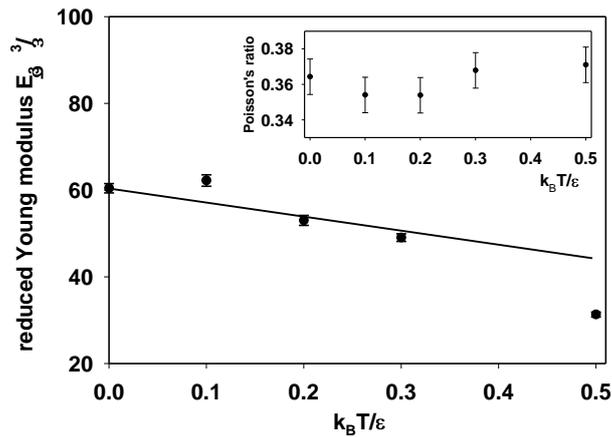

**Figure** 6: Symbols: reduced Young's modulus (main figure) and Poisson's ratio (inset) for the bulk fcc Lennard-Jones crystal as a function of the reduced temperature. The lines in the main figure and the inset are from ref [83]. The Young's modulus decreases with temperature; the Poisson's ratio is compatible with a constant within error bars.

| $k_B T/\varepsilon$ | $E\sigma^3/\varepsilon$ | $\nu$ |
|---|---|---|
| 0 | 60.5 | 0.364 |
| 0.1 | 62.2 | 0.354 |
| 0.2 | 53.0 | 0.354 |
| 0.3 | 49.0 | 0.368 |
| 0.5 | 31.1 | 0.371 |

**Table 1:** Reduced Young's modulus and Poisson's ratio for the bulk fcc Lennard-Jones crystal at different reduced temperatures (see text).

### III_2 Thin film

**Zero temperature calculations.** The two loading experiments (Figure 2) are performed on the five film thicknesses (4 to 12 layers). As for the bulk, the loading stress is chosen very small to be in the linear regime of deformations (no limitation due to thermal fluctuations). The results for the five (redundant) elastic quantities are shown in Figure 7. In all cases the data are shown versus the inverse film thickness expressed in terms of the number of planes. The bulk results, obtained with periodic boundary conditions are reported as if the number of planes was infinite. All curves are monotonous, showing smooth variations, with bulk results in agreement with the extrapolated evolution of the elastic constants to the infinite system. In particular, the anisotropy, revealed in the films by the differences between the three Poisson's ratio and the two Young's moduli, reduces when the film thickness increases, and is compatible with its vanishing for the infinite thickness. The consistency between the five mechanical constants is excellent whatever the system size: the parameter $\alpha = (E_1 \nu_{31})/(E_3 \nu_{13})$ (see inset in Figure 7) is compatible, within 2%, with the value of 1, expected for a perfectly symmetric matrix.

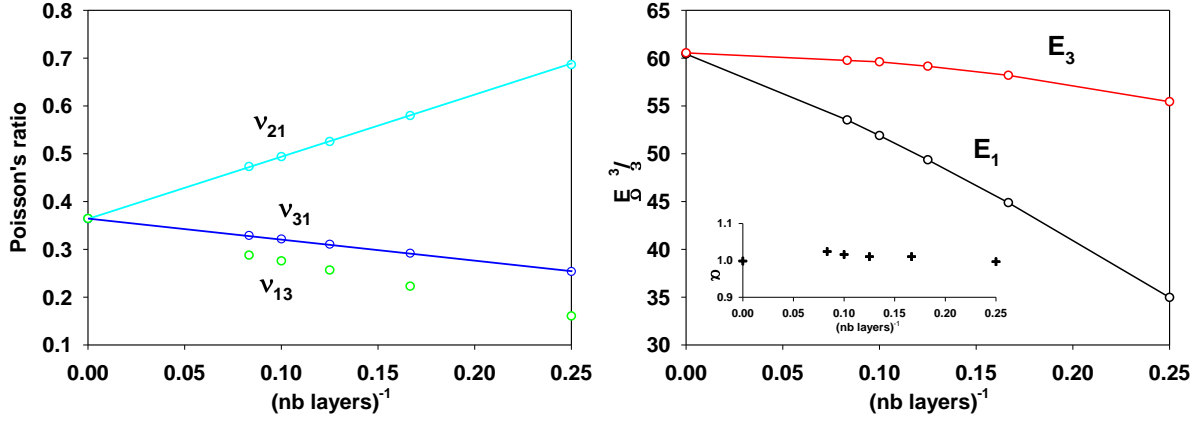

**Figure** 7: Poisson's ratio and Young's moduli at zero temperature for the fcc Lennard-Jones film, versus the inverse of its thickness given by the number of atomic layers. The lines for the Young's moduli are guides to the eye; the straight lines for the Poisson's ratio are issued from the bulk value (inverse thickness equal to zero) and best fit the data for finite thicknesses. Inset: evolution of the parameter $\alpha = (E_1 \nu_{31})/(E_3 \nu_{13})$ versus the film thickness.

The evolution of the three Poisson's ratio is almost linear with the inverse system size, even for the thinnest films. Departure from linearity is the largest for $\nu_{13}$ but never exceeds 5%. An interesting point is that their evolution with the film thickness is compatible with an essentially constant average value: $(\nu_{21} + \nu_{31} + \nu_{13})/3$ is independent of the film thickness. On the other hand, the evolution of the Young's moduli departs from a linear behavior for the smallest thicknesses. This non-linearity is also observed if one draws the compliance matrix coefficients (Eq 3): $1/E_1, -\nu_{21}/E_1$ etc. The general trend is a softening of the film when its thickness decreases, with amplification for the ultrathin films.

The stiffness matrix has been calculated by inversion of the numerical compliance matrix, without making any assumption on the symmetry. The five independent constants, $C_{11}$, $C_{33}$, $C_{12}$, $C_{13}$ and $C_{31}$ are shown in Figure 8 as a function of the inverse system thickness given by the number of layers. The symmetry of the stiffness matrix is well verified since $C_{13}=C_{31}$ within 5%. As previously, the anisotropy of the mechanical properties for the film appears through the differences between the two diagonal terms ($C_{11}$ and $C_{33}$) and the two off-diagonal terms ($C_{12}$ and $C_{13} = C_{31}$). As previously, the bulk values agree well with the expected extrapolation of the film constants for infinite thickness. The dependence with the inverse system size is close to linearity on the whole range of thicknesses for all stiffness constants. This is at odd with what was observed for the Young's moduli, which exhibited a poorly linear behavior with the inverse system size.

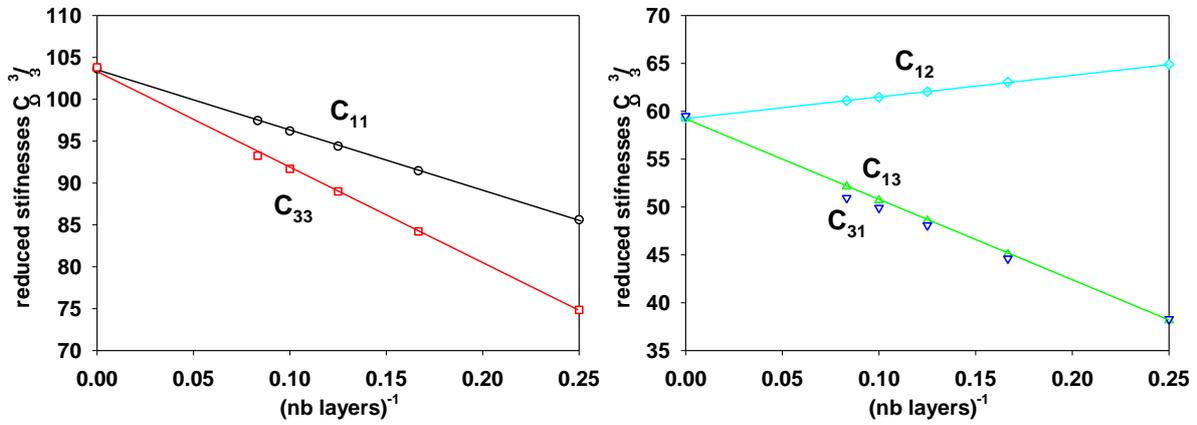

**Figure** 8: Stiffness matrix coefficients (eq 2) at zero temperature for the fcc Lennard-Jones film, versus the inverse of its thickness given by the number of atomic layers (eq 3). The straight lines are guides to the eye.

**Finite temperature calculations.** The temperature of the film has been varied between 0.1 and 0.5 in reduced units by increments of 0.1. The thermal agitation of the surface atoms increases significantly with temperature. At $T^* = 0.4$ and above, the thermal agitation is so large that the surface atoms are able to overcome the energy barrier that normally maintains them at their relative position in the crystal. The consequence is atomic migration on the external surface, with the formation of vacancies and adatoms. Since the film deforms irreversibly under loading (plasticity), the elastic constants become meaningless at high temperatures. Even when the structure of the film remains stable, the thermal fluctuations are large, and it is difficult to reach a good accuracy. In particular, at $T^* = 0.3$ and for the largest values of the stress used for the loadings, the film deformations are accompanied by atomic displacements. It was thus necessary to use smaller values of the stress, resulting in a lack of accuracy. For $T^* = 0.1$ and 0.2 the accuracy is comparable to that for the bulk solid.

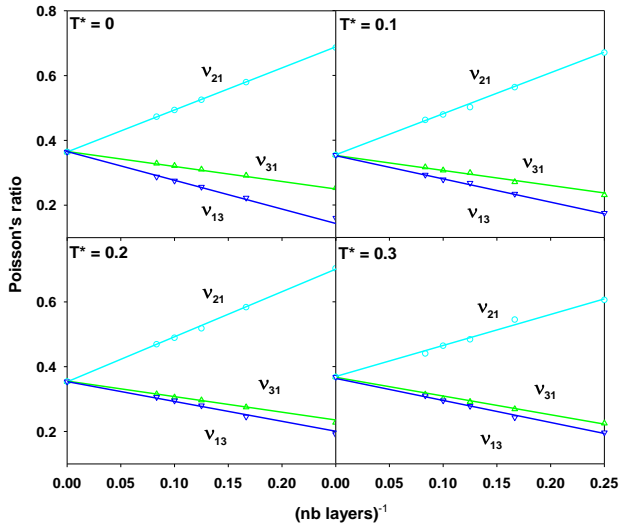

**Figure** 9: Poisson's ratio at different reduced temperatures (given in the panels) for the fcc Lennard-Jones films, versus the inverse of its thickness given by the number of atomic layers. The straight lines drawn in the figures are guides to the eye passing through the bulk value (inverse thickness equal to zero) and best fitting the data for finite thicknesses.

The results for the Poisson's ratio are given in Figure 9. As previously, the data are plotted as a function of the inverse film thickness given by the number of atomic layers. As can be seen, for all temperatures, the Poisson's ratios follow an essentially linear behavior with the inverse of the film thickness. Note that this linearity extends remarkably well even for the thinnest film. The largest departure from linearity never exceeds 5% whatever the temperature. Here again the evolution of the average value of the Poisson ratios $(\nu_{21}+\nu_{31}+\nu_{13})/3$ with the film thickness is compatible with a constant.

The Poisson ratios prove to be barely dependent on temperature. For instance, the largest variations in the bulk values never exceed 3%, and those for the thinnest film remain below 10% for $\nu_{21}$ and $\nu_{31}$ and below 15% for $\nu_{13}$. We however note the following trend: the differences between the three Poisson's ratios tend to decrease when $T$ increases. This is easy to observe for $\nu_{31}$ and $\nu_{13}$. The anisotropy of the thin film is thus attenuated by increasing the temperature.

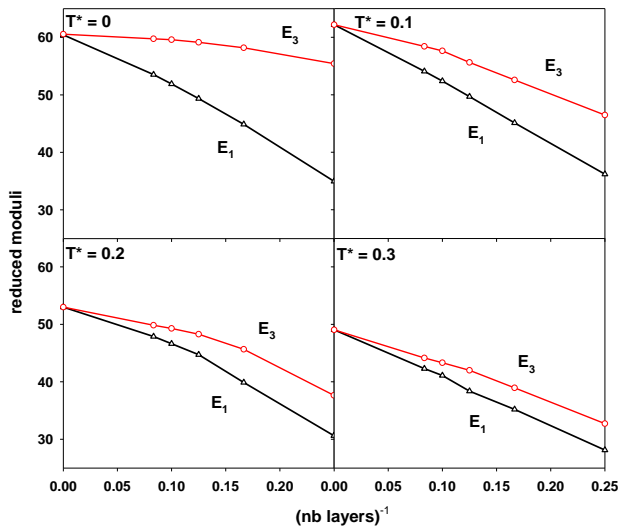

**Figure** 10: Young's moduli at different reduced temperatures (given in the panels) for the fcc Lennard-Jones films, versus the inverse of its thickness given by the number of atomic layers. The lines are guides to the eye.

The reduced Young's moduli obtained at different temperatures are given in Figure 10. For all temperatures, the general trend is a softening of the film when the thickness decreases, the dependence being non-linear with the inverse of the thickness. The data also show that the higher the temperature the lower the moduli, whatever the film thickness. In the in-plane directions, the decrease of $E_1$ reaches 20% between $T^* = 0$ and 0.3 for all thicknesses, while in the direction perpendicular to the film, the reduction of $E_3$ reaches 40% for the thinnest film (4 layers) between $T^* = 0$ and 0.3. The Young modulus in the out-of-plane direction is thus more affected by the temperature, probably in connection with the highest mobility of surface atoms.

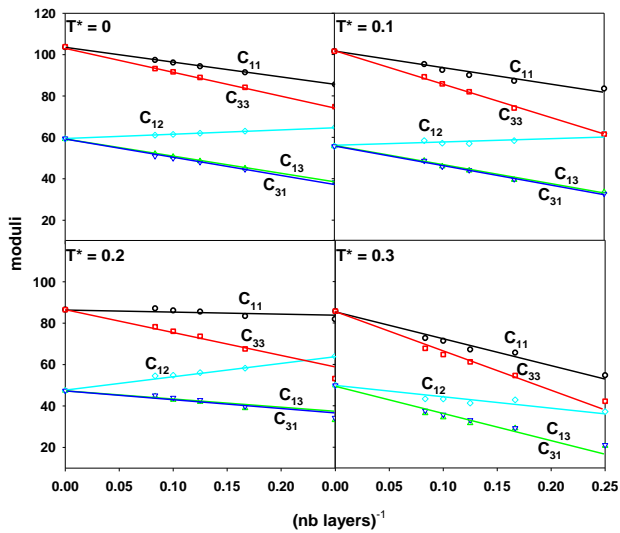

**Figure** 11: Stiffness matrix coefficients (Eq 2) at different reduced temperatures (given in the panels) for the fcc Lennard-Jones films, versus the inverse of its thickness given by the number of atomic layers. The straight lines are guides to the eye.

The Young moduli combined with the three Poisson's ratio give the compliance matrix (eq 3) that has been inverted to get the stiffness matrix (eq 2) for each temperature and film thickness. The five constants $C_{11}$, $C_{33}$, $C_{12}$, $C_{13}$ and $C_{31}$ are given in Figure 11. As expected from symmetry considerations of the matrix, the $C_{13}$ and $C_{31}$ constants are almost superimposed: the ratio between both remains equal to 1 within 2.5%. For all temperatures, the stiffness decreases for thinner films. As can be seen the stiffness constants exhibit a linear dependence with the inverse film thickness, which was not the case for the compliance matrix (see the discussion below). We also note an increase of the anisotropy (differences between the elastic constants) when the film thickness decreases.

A clear picture of the evolution of the elastic constants with the temperature does not emerge. The general trend is a decrease of the stiffness parameters when temperature increases, following essentially the behavior observed for the bulk solid. The evolution of the slope of the curves is weak below $T^* = 0.2$. At $T^* = 0.3$ the slope become significantly more negative for all elastic constants, showing a drastic decrease of the thin film stiffness. A careful analysis of the data shows however a non-monotonous dependence of the elastic constants with temperature, as was observed for the bulk crystal. A possible source of errors could be due to the high sensitivity of the compliance-to-stiffness inversion process to the original errors in the data. It has been tried to perform an inversion taking into account the symmetry constraint ($C_{13} = C_{31}$), but no significant improvement was observed. The errors in the data are most probably significantly larger than estimated. In particular, the Monte Carlo sampling may miss or overestimate some vibration modes in the simulation box associated with collective movements.[83]

**III_3 Discussion**

The most striking behavior shown by the data is the linear dependence of the stiffness constants with the inverse film thickness, from zero up to the highest temperature considered in this study ($T^* = 0.3$) above which the film loses its integrity under loading. The discrepancy never exceeds 5% at zero temperature (all thicknesses) or for thicknesses larger or equal to 6 layers (all temperatures), and remains below 15% in the worst case (four layers, high temperature).

The linear behavior of the elastic matrix versus the inverse system size suggests to interpret the evolution of the stiffness as due to a surface stress term. This can be seen by using the ansatz that any nanosystem = bulk + surface[30,70,88] and writing the total force that applies laterally to an idealized thin film made of a bulk-like core and an effective surface stress. It has been shown that generally any elastic property $D$ of such system would deviates from its bulk value $D^b$ according to[66]

$$D = D^b + d\frac{h^s}{h} \qquad (3)$$

where $h$ is the thickness of the nanostructure, $h^s$ corresponds to the length that sets the scale below which the surface elastic effects are significant, $d$ depends on the geometry, and $dh^s$ somehow relates to the surface elastic constants of the system. Applying this equation to the stiffness matrix, and replacing the thickness by the number of atomic planes, one gets:

$$C_{ij} = C_{ij}^b + \frac{c_{ij}^s}{N} \qquad . \qquad (4)$$

The $C_{ij}^s$ may be interpreted as the surface stiffness constants associated with the first atomic layer, while the $C_{ij}^b$ are the bulk values. These surface constants can be deduced from the best fit of the data shown in Figure 11 by extracting the slopes. The results are given in Table 2. This surface matrix is symmetric, with negative values except for $C_{12}^s$. The presence of the surfaces softens the thin film. An interesting point is that this surface matrix has non-negligible components in the z direction, perpendicular to the surface. This is actually unexpected in the usual formalism of surface stiffness. A possible interpretation is that the surface stress model applies only when the film thickness is large compared to the range of relaxation of the outermost layers. In our case, the film is so thin that the two external surfaces are not independent: the elastic constants in the direction perpendicular to the surface exhibit dependence with the film thickness. Another explanation could be found in the work of Chang et al.[99] who have proposed a lattice model where only the total strain energy density of an ultrathin film is taken into account, without any consideration of surface effect. They have observed a size dependence of the elastic constants, in particular for the out-of-plane stiffness coefficient $C_{13}$. This indicates that the surface effects might not be the only source of size dependence of the elastic constants of ultrathin films, yet the fundamental mechanisms are not fully understood. Further investigations are necessary to elucidate this point.

| T* | $C^b_{ii}$* | $C^b_{ij}$* | $C^s_{11}$* | $C^s_{33}$* | $C^s_{21}$* | $C^s_{31}$* |
|---|---|---|---|---|---|---|
| 0 | 103.7 | 59.4 | -71.8 | -118 | +21.6 | -83.6 |
| 0.1 | 101.6 | 55.7 | -81.6 | -159 | +20.4 | -94.8 |
| 0.2 | 86.5 | 47.4 | -11.2 | -110 | +63.6 | -39.6 |
| 0.3 | 85.8 | 49.9 | -132 | -192 | -58 | -136 |

**Table 2**: Reduced stiffness coefficients for the bulk and surface contributions (see eq 4) for an fcc Lennard-Jones thin film, at different reduced temperatures.

**IV Conclusion**

The simple Lennard-Jones crystal has been considered to model the elastic properties of thin films as a function of temperature and film thickness. The method consists in performing mechanical loadings and measuring the corresponding deformations. The anisotropy of the films requires at least four independent measurements (shear is not considered in this study). The calculations are performed in the isotherm-isostress ensemble by Monte Carlo simulations. The Young's moduli and Poisson's ratio appearing in the compliance matrix have been calculated between 0 and 0.5 in reduced units, for the bulk crystal and for the film of thickness 4 to 12 atomic layers. The bulk moduli decrease with temperature, while the Poisson's ratio remains essentially constant. For the films, one observes a softening of the solid when the thickness diminishes, in particular in the in-plane directions. The evolution with temperature shows a global softening with increasing T, the effect being the largest in the out-of-plane direction. At T* = 0.4 and above, the mobility of the surface atoms is too large, and the film deforms irreversibly under loading.

The coefficients of the stiffness matrix exhibit a linear dependence with the inverse of the system size. This has been analyzed in the framework of a bulk+surface contribution, and the corresponding surface stiffness has been determined from the corresponding slopes. However, the out-of-plane coefficients are non-zero, which is not compatible with the surface stress formalism. The bulk+surface decomposition is thus not justified in the case of very thin films, essentially because the surfaces are so close that they influence each other (the distance associated to surface relaxations exceeds the film thickness). However, the dependence of the stiffness matrix with the inverse thickness is remarkably linear and deserves further studies.

**Acknowledgement**

The author acknowledges financial support from ANR-17-CE30-002-04.

**References**

(1) Landau, L. D.; Lifshitz, E. M. *Theory of Elasticity*; Pergamon Press: London, 1959.

(2) Meehan, F. T. The Expansion of Charcoal on Sorption of Carbon Dioxide. *Proc. R. Soc. London A* **1927**, *115*, 199-207.

(3) Bangham, D. H.; Fakhoury, N. The Expansion of Charcoal Accompanying Sorption of Gases and Vapours. *Nature* **1928**, *122*, 681-682.

(4) Dušek, K. Responsive Gels: Volume Transitions I. *Adv. Polym. Sci.* **1993**, *109*, 1-267.

(5) Scherer, G. W.; Smith, D. M.; Stein, D. Deformation of Aerogels During Characterization. *J. Non-Cryst. Solids* **1995**, *186*, 309-315.

(6) Ogieglo, W.; Wormeester, H.; Eichhorn, K.-J.; Wessling, M.; Benes, N. E. In Situ Ellipsometry Studies on Swelling of Thin Polymer Films: A Review. *Prog. Polym. Sci.* **2015**, *42*, 42-78.

(7) Kulasinski, K.; Guyer, R.; Derome, D.; Carmeliet, J. Poroelastic Model for Adsorption-Induced Deformation of Biopolymers Obtained from Molecular Simulations. *Phys. Rev. E* **2015**, *92*, 022605.

(8) Scherer, G. W. Dilatation of Porous Glass. *J. Am. Ceram. Soc.* **1986**, *69*, 473-480.

(9) Yates, D. J. C. The Expansion of Porous Glass on the Adsorption of Non-Polar Gases. *Proc. R. Soc. London A* **1954**, *224*, 526-544.

(10) Haines, R. S.; McIntosh, R. Length Changes of Activated Carbon Rods Caused by Adsorption of Vapors. *J. Chem. Phys.* **1947**, *15*, 28-38.

(11) Balzer, C.; Braxmeier, S.; Neimark, A. V.; Reichenauer, G. Deformation of Microporous Carbon During Adsorption of Nitrogen, Argon, Carbon Dioxide, and Water Studied by in Situ Dilatometry. *Langmuir* **2015**, *31*, 12512-12519.

(12) Fomkin, A. A.; Shkolin, A. V.; Pulin, A. L.; Men'shchikov, I. E.; Khozina, E. V. Adsorption-Induced Deformation of Adsorbents. *Colloid Journal* **2018**, *80*, 578-586.

(13) Biot, M. A. General Theory of Three-Dimensional Consolidation. *J. Appl. Phys.* **1941**, *12*, 155-164.

(14) Coussy, O. *Poromechanics*; John Wiley & Sons: New York, 2004.

(15) Coasne, B.; Weigel, C.; Polian, A.; Kint, M.; Rouquette, J.; Haines, J.; Foret, M.; Vacher, R.; Rufflé, B. Poroelastic Theory Applied to the Adsorption-Induced Deformation of Vitreous Silica. *J. Phys. Chem. B* **2014**, *118*, 14519-14525.

(16) Brochard, L.; Vandamme, M.; Pellenq, R. J. M. Poromechanics of Microporous Media. *J. Mech. Phys. Solids* **2012**, *60*, 606-622.

(17) Gor, G. Y.; Huber, P.; Bernstein, N. Adsorption-Induced Deformation of Nanoporous Materials—a Review. *Appl. Phys. Rev.* **2017**, *4*, 011303.

(18) Vandamme, M.; Brochard, L.; Lecampion, B.; Coussy, O. Adsorption and Strain: The Co2-Induced Swelling of Coal. *J. Mech. Phys. Solids* **2010**, *58*, 1489-1505.

(19) Vandamme, M. Coupling between Adsorption and Mechanics (and Vice Versa). *Curr. Opin. Chem. eng.* **2019**, *24*, 12-18.

(20) Dolino, G.; Bellet, D.; Faivre, C. Adsorption Strains in Porous Silicon. *Phys. Rev. B* **1996**, *54*, 17919-17929.

(21) Boissiere, C.; Grosso, D.; Lepoutre, S.; Nicole, L.; Bruneau, A. B.; Sanchez, C. Porosity and Mechanical Properties of Mesoporous Thin Films Assessed by Environmental Ellipsometric Porosimetry. *Langmuir* **2005**, *21*, 12362-12371.

(22) Herman, T.; Day, J.; Beamish, J. Deformation of Silica Aerogel During Fluid Adsorption. *Phys. Rev. B* **2006**, *73*, 094127.

(23) Dourdain, S.; Britton, D. T.; Reichert, H.; Gibaud, A. Determination of the Elastic Modulus of Mesoporous Silica Thin Films by X-Ray Reflectivity Via the Capillary Condensation of Water. *Appl. Phys. Lett.* **2008**, *93*, 183108.

(24) Prass, J.; Müter, D.; Fratzl, P.; Paris, O. Capillarity-Driven Deformation of Ordered Nanoporous Silica. *Appl. Phys. Lett.* **2009**, *95*, 083121.

(25) Schoen, M.; Paris, O.; Günther, G.; Müter, D.; Prass, J.; Fratzl, P. Pore-Lattice Deformations in Ordered Mesoporous Matrices: Experimental Studies and Theoretical Analysis. *Phys. Chem. Chem. Phys.* **2010**, *12*, 11267-11279.

(26) Sharifi, P.; Marmiroli, B.; Sartori, B.; Cacho-Nerin, F.; Keckes, J.; Amenitsch, H.; Paris, O. Humidity-Driven Deformation of Ordered Mesoporous Silica Films. *Bioinspired, Biomimetic Nanobiomater.* **2014**, *3*, 183-190.

(27) Grosman, A.; Puibasset, J.; Rolley, E. Adsorption-Induced Strain of a Nanoscale Silicon Honeycomb. *Europhys. Lett.* **2015**, *109*, 56002.

(28) Liu, M.; Wu, J.; Gan, Y.; Chen, C. Q. The Pore-Load Modulus of Ordered Nanoporous Materials with Surface Effects. *AIP Adv.* **2016**, *6*, 035324.

(29) Rolley, E.; Garroum, N.; Grosman, A. Using Capillary Forces to Determine the Elastic Properties of Mesoporous Materials. *Phys. Rev. B* **2017**, *95*, 064106.

(30) Zhang, Y.; Zhuo, L. J.; Zhao, H. S. Determining the Effects of Surface Elasticity and Surface Stress by Measuring the Shifts of Resonant Frequencies. *Proc. R. Soc. London A* **2013**, *469*, 20130449.

(31) Fu, W.-E.; Chang, Y.-Q.; He, B.-C.; Wu, C.-L. Determination of Young's Modulus and Poisson's Ratio of Thin Films by X-Ray Methods. *Thin Solid Films* **2013**, *544*, 201-205.

(32)     Hoogeboom-Pot, K. M.; Turgut, E.; Hernandez-Charpak, J. N.; Shaw, J. M.; Kapteyn, H. C.; Murnane, M. M.; Nardi, D. Nondestructive Measurement of the Evolution of Layer-Specific Mechanical Properties in Sub-10 Nm Bilayer Films. *Nano Lett.* **2016**, *16*, 4773-4778.

(33)     Puibasset, J. Adsorption-Induced Deformation of a Nanoporous Material: Influence of the Fluid–Adsorbent Interaction and Surface Freezing on the Pore-Load Modulus Measurement. *J. Phys. Chem. C* **2017**, *121*, 18779-18788.

(34)     Cohan, L. H. Sorption Hysteresis and the Vapor Pressure of Concave Surfaces. *J. Am. Chem. Soc.* **1938**, *60*, 433-435.

(35)     Everett, D. H.; Whitton, W. I. A General Approach to Hysteresis. *Trans. Faraday Soc.* **1952**, *48*, 749.

(36)     Everett, D. H.; Smith, F. W. A General Approach to Hysteresis. Part2: Development of the Domain Theory. *Trans. Faraday Soc.* **1954**, *50*, 187.

(37)     Mason, G. The Effect of Pore Space Connectivity on the Hysteresis of Capillary Condensation in Adsorption Desorption Isotherms. *J. Colloid Interface Sci.* **1982**, *88*, 36-46.

(38)     Evans, R.; Marini Bettolo Marconi, U.; Tarazona, P. Capillary Condensation and Adsorption in Cylindrical and Slit - Like Pores. *J. Chem. Soc., Faraday Trans. 2* **1986**, *82*, 1763-1787.

(39)     Kierlik, E.; Rosinberg, M. L.; Tarjus, G.; Monson, P. A. Phase Diagrams of Single-Component Fluids in Disordered Porous Materials: Predictions from Integral-Equation Theory. *J. Chem. Phys.* **1997**, *106*, 264-279.

(40)     Kierlik, E.; Rosinberg, M. L.; Tarjus, G.; Viot, P. Equilibrium and out-of-Equilibrium (Hysteretic) Behavior of Fluids in Disordered Porous Materials : Theoretical Predictions. *Phys. Chem. Chem. Phys.* **2001**, *3*, 1201-1206.

(41)     Kierlik, E.; Monson, P. A.; Rosinberg, M. L.; Tarjus, G. Adsorption Hysteresis and Capillary Condensation in Disordered Porous Solids: A Density Functional Study. *J. Phys.: Condens. Matter* **2002**, *14*, 9295-9315.

(42)     Puibasset, J. Adsorption/Desorption Hysteresis of Simple Fluids Confined in Realistic Heterogeneous Silica Mesopores of Micrometric Length: A New Analysis Exploiting a Multiscale Monte Carlo Approach. *J. Chem. Phys.* **2007**, *127*, 154701.

(43)     Bonnet, F.; Lambert, T.; Cross, B.; Guyon, L.; Despetis, F.; Puech, L.; Wolf, P. E. Evidence for a Disorder-Driven Phase Transition in the Condensation of 4he in Aerogels. *Europhys. Lett.* **2008**, *82*, 56003.

(44)     Puibasset, J. Monte Carlo Multiscale Simulation Study of Argon Adsorption/Desorption Hsyteresis in Mesoporous Heterogeneous Tubular Pores Like Mcm-41 or Oxidized Porous Silicon. *Langmuir* **2009**, *25*, 903-911.

(45)     Puibasset, J. Counting Metastable States within the Adsorption/Desorption Hysteresis Loop: A Molecular Simulation Study of Confinement in Heterogeneous Pores. *J. Chem. Phys.* **2010**, *133*, 104701.

(46)     Puibasset, J. Numerical Characterization of the Density of Metastable States within the Hysteresis Loop in Disordered Systems. *J. Phys.: Condens. Matter* **2011**, *23*, 035106.

(47)     Puibasset, J. Stability Intervals of Metastable States in Hysteretic Systems. *Phys. Rev. E* **2011**, *84*, 061126.

(48)     Puibasset, J. Fluid Adsorption in Linear Pores: A Molecular Simulation Study of the Influence of Heterogeneities on the Hysteresis Loop and the Distribution of Metastable States. *Mol. Simul.* **2014**, *40*, 690-697.

(49)     Bonnet, F.; Melich, M.; Puech, L.; Angles d'Auriac, J.-C.; Wolf, P.-E. Deciphering the Mechanisms of Condensation and Evaporation in a Complex Porous Material. *Langmuir* **2019**, *submitted*.

(50)     Amberg, C. H.; McIntosh, R. A Study of Adsorption Hysteresis by Means of Length Changes of a Rod of Porous Glass. *Can. J. Chem.* **1952**, *30*, 1012-1032.

(51)     Quinn, H. W.; McIntosh, R. The Hysteresis Loop in Adsorption Isotherms on Porous Vycor Glass and Associated Dimensional Changes of the Adsorbent. Ii. *Can. J. Chem.* **1957**, *35*, 745-756.

(52)     Fomkin, A. A. Adsorption of Gases, Vapors and Liquids by Microporous Adsorbents. *Adsorption* **2005**, *11*, 425-436.

(53)     Günther, G.; Prass, J.; Paris, O.; Schoen, M. Novel Insights into Nanopore Deformation Caused by Capillary Condensation. *Phys. Rev. Lett.* **2008**, *101*, 086104.

(54)     Grosman, A.; Ortega, C. Influence of Elastic Deformation of Porous Materials in Adsorption-Desorption Process: A Thermodynamic Approach. *Phys. Rev. B* **2008**, *78*, 085433.

(55)     Grosman, A.; Ortega, C. Influence of Elastic Strains on the Adsorption Process in Porous Materials: An Experimental Approach. *Langmuir* **2009**, *25*, 8083-8093.

(56)     Grosman, A.; Ortega, C. Influence of Elastic Strains on the Adsorption Process in Porous Materials. Thermodynamics and Experiments. *Appl. Surf. Sci.* **2010**, *256*, 5210-5215.

(57)     Grosman, A.; Ortega, C. Cavitation in Metastable Fluids Confined to Linear Mesopores. *Langmuir* **2011**, *27*, 2364-2374.

(58)     Ekinci, K. L.; Roukes, M. L. Nanoelectromechanical Systems. *Rev. Sci. Instrum.* **2005**, *76*, 061101.

(59) Li, X.; Ono, T.; Wang, Y.; Esashi, M. Ultrathin Single-Crystalline-Silicon Cantilever Resonators: Fabrication Technology and Significant Specimen Size Effect on Young's Modulus. *Appl. Phys. Lett.* **2003**, *83*, 3081-3083.

(60) Park, S. H.; Kim, J. S.; Park, J. H.; Lee, J. S.; Choi, Y. K.; Kwon, O. M. Molecular Dynamics Study on Size-Dependent Elastic Properties of Silicon Nanocantilevers. *Thin Solid Films* **2005**, *492*, 285-289.

(61) Gordon, M. J.; Baron, T.; Dhalluin, F.; Gentile, P.; Ferret, P. Size Effects in Mechanical Deformation and Fracture of Cantilevered Silicon Nanowires. *Nano Lett.* **2009**, *9*, 525-529.

(62) Sadeghian, H.; Goosen, H.; Bossche, A.; Thijsse, B.; van Keulen, F. On the Size-Dependent Elasticity of Silicon Nanocantilevers: Impact of Defects. *J. Phys. D: Appl. Phys.* **2011**, *44*, 072001.

(63) Gumbsch, P.; Daw, M. S. Interface Stresses and Their Effects on the Elastic Moduli of Metallic Multilayers. *Phys. Rev. B* **1991**, *44*, 3934-3938.

(64) Wolf, D. Surface-Stress-Induced Structure and Elastic Behavior of Thin Films. *Appl. Phys. Lett.* **1991**, *58*, 2081-2083.

(65) Cammarata, R. C. Surface and Interface Stress Effects in Thin Films. *Prog. Surf. Sci.* **1994**, *46*, 1-38.

(66) Miller, R. E.; Shenoy, V. B. Size-Dependent Elastic Properties of Nanosized Structural Elements. *Nanotechnology* **2000**, *11*, 139-147.

(67) Fedorchenko, A. I.; Wang, A.-B.; Cheng, H. H. Thickness Dependence of Nanofilm Elastic Modulus. *Appl. Phys. Lett.* **2009**, *94*, 152111.

(68) Haiss, W. Surface Stress of Clean and Adsorbate-Covered Solids. *Rep. Prog. Phys.* **2001**, *64*, 591.

(69)	Van Workum, K.; de Pablo, J. J. Local Elastic Constants in Thin Films of an Fcc Crystal. *Phys. Rev. E* **2003**, *67*, 031601.

(70)	Müller, P.; Saúl, A. Elastic Effects on Surface Physics. *Surf. Sci. Rep.* **2004**, *54*, 157-258.

(71)	Zhou, L. G.; Huang, H. Are Surfaces Elastically Softer or Stiffer? *Appl. Phys. Lett.* **2004**, *84*, 1940-1942.

(72)	Müller, P.; Saùl, A.; Leroy, F. Simple Views on Surface Stress and Surface Energy Concepts. *Adv. Nat. Sci.: Nanosci. Nanotechnol.* **2014**, *5*, 013002.

(73)	Domb, C.; Salter, L. Cix. The Zero Point Energy and Θ Crystals. *Philos. Mag.* **1952**, *43*, 1083-1089.

(74)	Horton, G. K. Ideal Rare-Gas Crystals. *Am. J. Phys.* **1968**, *36*, 93-119.

(75)	Klein, M. L.; Hoover, W. G. Comparison of Classical Monte Carlo Experiments with Improved Self-Consistent Phonon Theory: Thermodynamic Properties of Solid Xe. *Phys. Rev. B* **1971**, *4*, 539-542.

(76)	De Wette, F. W.; Fowler, L. H.; Nijboer, B. R. A. Lattice Dynamics, Thermal Expansion and Specific Heat of a Lennard-Jones Solid in the Quasi-Harmonic Approximation. *Physica* **1971**, *54*, 292-304.

(77)	Westera, K.; Cowley, E. R. Cell-Cluster Expansion for an Anharmonic Solid. *Phys. Rev. B* **1975**, *11*, 4008-4016.

(78)	Mohazzabi, P.; Behroozi, F. Simple Classical Calculation of Thermal Expansion for Rare-Gas Solids. *Phys. Rev. B* **1987**, *36*, 9820-9823.

(79)	Tretiakov, K. V.; Wojciechowski, K. W. Quick and Accurate Estimation of the Elastic Constants Using the Minimum Image Method. *Comput. Phys. Comm.* **2015**, *189*, 77-83.

(80) Phuong, D. D.; Hoa, N. T.; Van Hung, V.; Khoa, D. Q.; Hieu, H. K. Mechanical Properties of Metallic Thin Films: Theoretical Approach. *Eur. Phys. J. B* **2016**, *89*, 84.

(81) Squire, D. R.; Holt, A. C.; Hoover, W. G. Isothermal Elastic Constants for Argon. Theory and Monte Carlo Calculations. *Physica* **1969**, *42*, 388-397.

(82) Sprik, M.; Impey, R. W.; Klein, M. L. Second-Order Elastic Constants for the Lennard-Jones Solid. *Phys. Rev. B* **1984**, *29*, 4368-4374.

(83) Quesnel, D. J.; Rimai, D. S.; DeMejo, L. P. Elastic Compliances and Stiffnesses of the Fcc Lennard-Jones Solid. *Phys. Rev. B* **1993**, *48*, 6795-6807.

(84) Ray, J. R. Fluctuations and Thermodynamic Properties of Anisotropic Solids. *J. Appl. Phys.* **1982**, *53*, 6441-6443.

(85) Ray, J. R. Elastic Constants and Statistical Ensembles in Molecular Dynamics. *Comput. Phys. Rep.* **1988**, *8*, 109-151.

(86) Miyazaki, N.; Shiozaki, Y. Calculation of Mechanical Properties of Solids Using Molecular Dynamics Method. *JSME Int. J. ser A* **1996**, *39*, 606-612.

(87) Van Workum, K. L. Mechanical Properties of Solids and Nanoscopic Structures Via Molecular Simulations. Ph. D. Thesis, University of Wisconsin-Madison, 2002.

(88) Shenoy, V. B. Atomistic Calculations of Elastic Properties of Metallic Fcc Crystal Surfaces. *Phys. Rev. B* **2005**, *71*, 094104.

(89) Clavier, G.; Desbiens, N.; Bourasseau, E.; Lachet, V.; Brusselle-Dupend, N.; Rousseau, B. Computation of Elastic Constants of Solids Using Molecular Simulation: Comparison of Constant Volume and Constant Pressure Ensemble Methods. *Mol. Simul.* **2017**, *43*, 1413-1422.

(90) Cowley, E. R. Some Monte Carlo Calculations for the Lennard-Jones Solid. *Phys. Rev. B* **1983**, *28*, 3160-3163.

(91) Li, M.; Johnson, W. L. Fluctuations and Thermodynamic Response Functions in a Lennard-Jones Solid. *Phys. Rev. B* **1992**, *46*, 5237-5241.

(92) Quesnel, D. J.; Rimai, D. S.; DeMejo, L. P. The Poisson Ratio for an Fcc Lennard-Jones Solid. *Solid State Commun.* **1993**, *85*, 171-175.

(93) Iwaki, T. Molecular Dynamics Study on Stress-Strain in Very Thin Film : Size and Location of Region for Defining Stress and Strain. *JSME Int. J. Ser. A* **1996**, *39*, 346-353.

(94) Heinz, H.; Vaia, R. A.; Farmer, B. L.; Naik, R. R. Accurate Simulation of Surfaces and Interfaces of Face-Centered Cubic Metals Using 12−6 and 9−6 Lennard-Jones Potentials. *J. Phys. Chem. C* **2008**, *112*, 17281-17290.

(95) Hamilton, J. C.; Wolfer, W. G. Theories of Surface Elasticity for Nanoscale Objects. *Surf. Sci.* **2009**, *603*, 1284-1291.

(96) Zhou, X.; Ren, H.; Huang, B.; Zhang, T. Size-Dependent Elastic Properties of Thin Films: Surface Anisotropy and Surface Bonding. *Sci. China Technol. Sc.* **2014**, *57*, 680-691.

(97) Streitz, F. H.; Cammarata, R. C.; Sieradzki, K. Surface-Stress Effects on Elastic Properties. I. Thin Metal Films. *Phys. Rev. B* **1994**, *49*, 10699-10706.

(98) Lu, P.; He, L. H.; Lee, H. P.; Lu, C. Thin Plate Theory Including Surface Effects. *Int. J. Solids Struct.* **2006**, *43*, 4631-4647.

(99) Chang, I. L.; Chang, S.-H.; Huang, J.-C. The Theoretical Model of Fcc Ultrathin Film. *Int. J. Solids Struct.* **2007**, *44*, 5818-5828.

(100) Broughton, J. Q.; Gilmer, G. H. Molecular Dynamics Investigation of the Crystal–Fluid Interface. Ii. Structures of the Fcc (111), (100), and (110) Crystal–Vapor Systems. *J. Chem. Phys.* **1983**, *79*, 5105-5118.

(101) Wolf, D.; Lutsko, J. F. Structurally Induced Supermodulus Effect in Superlattices. *Phys. Rev. Lett.* **1988**, *60*, 1170-1173.

(102) Cammarata, R. C.; Sieradzki, K. Effects of Surface Stress on the Elastic Moduli of Thin Films and Superlattices. *Phys. Rev. Lett.* **1989**, *62*, 2005-2008.

(103) Streitz, F. H.; Cammarata, R. C.; Sieradzki, K. Surface-Stress Effects on Elastic Properties. Ii. Metallic Multilayers. *Phys. Rev. B* **1994**, *49*, 10707-10716.

(104) Tęcza, G. W. Structural and Elastic Properties of Fcc/Fcc Metallic Multilayers: A Molecular-Dynamics Study. *Phys. Rev. B* **1992**, *46*, 15447-15451.

(105) Streitz, F. H.; Sieradzki, K.; Cammarata, R. C. Elastic Properties of Thin Fcc Films. *Phys. Rev. B* **1990**, *41*, 12285-12287.

(106) Wolf, D. Computer Simulation of Elastic and Structural Properties of Thin Films. *Surf. Sci.* **1990**, *225*, 117-129.

(107) Sun, C. T.; Zhang, H. Size-Dependent Elastic Moduli of Platelike Nanomaterials. *J. Appl. Phys.* **2003**, *93*, 1212-1218.

(108) Liang, L.; Li, M.; Qin, F.; Wei, Y. Temperature Effect on Elastic Modulus of Thin Films and Nanocrystals. *Philos. Mag.* **2013**, *93*, 574-583.

(109) Allen, M. P.; Tildesley, D. J. *Computer Simulation of Liquids*; Clarendon Press: Oxford, 1987.